\documentclass[prl,aps,superscriptaddress,groupedaddress, twocolumn]{revtex4}

\usepackage{xspace,amsmath,amsfonts,amsthm,amssymb,amsbsy}
\usepackage{graphicx}
\usepackage{amssymb}
\usepackage{color}
\usepackage[usenames,dvipsnames]{xcolor}
\usepackage[colorlinks=true,citecolor=Cerulean,linkcolor=RubineRed,urlcolor=Cerulean]{hyperref}

\definecolor{pink}{rgb}{1,0.078,0.57}

\definecolor{green}{rgb}{0,0.7,0.9}

\newcommand{\ket}[1] {\left\vert #1 \right\rangle}

\newcommand{\bra}[1] {\left\langle #1 \right|}
\newcommand{\braket}[2] {\langle #1 | #2 \rangle}

\newcommand{\dg}{^{\dagger}}

\graphicspath{{../images/}}

\begin{document}

\title{Two-level System coupled to Phonons: Full Analytical Solution }

\author{Aur\'elia Chenu}
\affiliation{Donostia International Physics Center,  E-20018 San Sebasti\'an, Spain}
\affiliation{IKERBASQUE, Basque Foundation for Science, E-48013 Bilbao, Spain}
\affiliation{Theoretical Division, Los Alamos National Laboratory, Los Alamos, New Mexico 87545, USA}

\author{Shiue-Yuan Shiau} 
\affiliation{Physics Division, National Center for Theoretical Sciences, Hsinchu 30013, Taiwan, Republic of China}

\author{Monique Combescot}
\affiliation{Sorbonne Universit\'e, CNRS, Institut des NanoSciences de Paris, 75005-Paris, France}


\begin{abstract}
We propose an analytical procedure to fully solve a two-level system coupled to phonons. 
 Instead of using the common formulation in terms of  linear and quadratic  system-phonon couplings, we  introduce different phonons depending on the system electronic level. We use this approach to recover known results for  the linear-coupling limit in a simple way. More importantly, we derive  results for the quadratic coupling induced by a phonon frequency change, a problem considered up to now as not analytically solvable. 
\end{abstract}

\maketitle

Coupling between a system and its environment is of primordial importance in science and emerging technologies, but its description at the microscopic level is extremely complicated \cite{BreuerBook,WeissBook, MayKuehnBook,ValkunasMancalBook,FerryBook,WisemanBook,SchlosshauerBook}.
Any consistent study of a quantum system  calls for an open description which includes the large, often poorly known environment that interacts with it.

A paradigmatic example of open systems is the `spin-boson' model \cite{WeissBook,Leggett1987a}, that describes a  two-level system coupled to a vibrational environment. 
In its simpler version, known as the `independent boson' model  \cite{MahanBook}, the system excited level is suddenly populated or depopulated through its coupling to a photon field.  While very simple, this model is not trivial  and already allows studying a  variety of phenomena that 
include spectral lineshapes \cite{MahanBook}, electron transfer \cite{Cai1989a,Morillo1996a}, electron-phonon interaction in quantum dots \cite{Vagov2002a, Hohenester2004a}, and quantum control \cite{Jirari2007a,Duke1965a, Steinbach1999a, Castella1999a, Axt1999a,Krummheuer2002a, Besombes2001a,ScullyBook, LoudonBook,BreuerBook}. 

The independent boson model fundamentally deals with the consequences of electronic excitations that induce a spatial shift in the vibrational modes  \cite{Silbey1980a, MukamelBook,NitzanBook}, and in some materials like aromatic hydrocarbons, a frequency change \cite{Munn1970a,Coropceanu2007a,Sluch2001a,Coropceanu2002a,Liu2007a,Uskov2000a,Muljarov2004a}. 
 Instead of  using the physically relevant phonons that depend on the occupied level of the electronic system, i.e. a diagonal representation, the common approach to this problem resorts to only using ground-phonons---the vibrations physically relevant when the system is in its ground level---%
even when the system is in its excited level. This off-diagonal representation gives rise to a linear coupling associated with the atom (or molecule) spatial shift and a quadratic coupling associated with the phonon frequency change. These couplings are then commonly eliminated using a polaron transformation  \cite{MahanBook}.
 Since then and until now, a plethora of studies follows this polaron procedure to address a diversity of problems, including `open systems', which presently is a very active field. 

 Although the polaron transformation can formally eliminate both linear and quadratic couplings, 
 it involves  calculations so tedious that to date, 
  analytical results have  been found for the linear limit only. 
  This representation leads to the idea that after its sudden excitation, the electronic system is dressed by a cloud of ground-phonons---whereas our treatment suggests to interpret the vibrations as dependent on the system excitation.
  %
  Approaches relying on cumulant expansion and diagrammatic Green's functions have also been used to study the broadening of the zero-phonon line when the quadratic coupling is included. However, these results were obtained through a weak-coupling expansion \cite{McCumber1963a, Krivoglaz1964a, Jones1978a, DiBartolo2010a}, through a long-time expansion \cite{Levenson1971a, Osadko1973a, Goupalov2002a, Hsu1984a, Hizhnyakov2002a}, or numerically \cite{Muljarov2004a}.

In this Letter, we propose a totally different procedure that relies on level-dependent phonons: The ground-phonons and the excited-phonons that describe the vibrational environment when the electronic system is, respectively, in its ground and excited level. 
 A compelling evidence demonstrating the strength of this alternative approach is that (i) all the results  for the linear limit become easy to derive, and more importantly, (ii) the same straightforward algebras make possible the handling of the quadratic coupling induced by a phonon frequency difference, a configuration considered up to now as not analytically solvable.

The hidden difficulty incurred by considering different phonon frequencies comes from the fact that the destruction of an excited-phonon not only corresponds to the destruction  but also the creation of a ground-phonon, making the phonon subspace associated with the excited level infinitely large when written in terms of ground-phonons. We will show that this difficulty is simply accounted for in the excited-phonon vacuum.  

\noindent\emph{\textbf{The two relevant phonon bases}}.--- 
We consider a two-level system coupled to phonons, which for simplicity are taken as one-dimensional harmonic oscillators. When the electronic system is in its ground level $\ket{g}$ with energy $\epsilon_g$, the Hamiltonian reads in first quantization  as
\begin{eqnarray}
\mathcal{H}_g =\Big( \epsilon_g + \frac{p_x^2}{2m}  + \frac{m \omega_g^2}{2} x^2\Big) \ket{g} \bra{g}, 
\end{eqnarray}
where $m$ and $\omega_g$ are the mass and frequency of the oscillator. By introducing the ground-phonon destruction operator 
\begin{eqnarray}  \label{3}
b_g= \sqrt{\frac{m\omega_g}{2\hbar}}\Big(\hat{x} +i    \frac{\hat{p}_x}{m\omega_g} \Big ),
\end{eqnarray} 
that fulfills $[b_g,b_g\dg]=1$, we obtain the well-known second-quantization form 
\begin{eqnarray} \label{Hg}
H_{g} &=&  \Big(\epsilon_g + \hbar \omega_g \big(b_g\dg b_g+1/2\big) \Big)a\dg_g a_g, 
\end{eqnarray}
with $a\dg_g \ket{v}= \ket{g}$ where $\ket{v}$ is the electronic system vacuum. 
The Hamiltonian eigenset reads  
\begin{eqnarray}
\ket{\psi_{g,p}} = \ket{g} \otimes \ket{p_g}, \, \, \, \, \,\, \, \, \, \,E_{g,p} = \epsilon_g +  \hbar \omega_g \left(p+1/2 \right)  , 
\end{eqnarray}
 where $\ket{p_g} = (1/\sqrt{p!})\,\,b_g^{\dagger \, p} \ket{0_g}$ contains $\bra{p_g}b_g\dg b_g  \ket{p_g}  =p$ ground-phonons, the state $\ket{0_g}$ being the ground-phonon vacuum defined as $b_g \ket{0_g} = 0$. \

 When the system is in its excited level $\ket{e}$ with energy $\epsilon_e$, 
the Hamiltonian is the same, except for a spatial shift $l$ related to the dipolar nature of the excited level, and a possibly different frequency, which, as shown below, induces a quadratic coupling mostly neglected in the absence of known procedure to handle it, 
\begin{eqnarray}
\mathcal{H}_e =\Big( \epsilon_e + \frac{p_x^2}{2m}  + \frac{m \omega_e^2}{2} (x-l)^2\Big) \ket{e} \bra{e}.
\end{eqnarray}
 The relevant `excited-phonon' operator then reads
\begin{eqnarray} \label{b}
b_e= \sqrt{\frac{m\omega_e}{2\hbar}}\Big((\hat{x} -l)+i    \frac{\hat{p}_{x}}{m\omega_e}\Big  ) , 
\end{eqnarray} 
that  also fulfills $[b_e,b_e\dg]=1$. This gives the $\mathcal{H}_e $ Hamiltonian in second quantization as 
\begin{eqnarray} \label{He}
H_{e} &=&  \Big(\epsilon_e + \hbar \omega_e \big(b_e\dg b_e+1/2\big) \Big)a\dg_e a_e, 
\end{eqnarray}
with $a\dg_e \ket{v} = \ket{e}$, 
its eigenset being 
 \begin{eqnarray} \label{eq:psi_e}
\ket{\psi_{e,p}} = \ket{e} \otimes \ket{p_e}, \, \, \, \, \,\, \, \, \, \,E_{e,p} = \epsilon_e + \hbar  \omega_e \left(p+1/2\right) , 
\end{eqnarray}
where $\ket{p_e} =
(1/\sqrt{p!})
 \, \, b_e^{\dagger \, p}\ket{0_e}$ contains $\bra{p_e} b_e\dg b_e  \ket{p_e}   =p$  
excited-phonons provided that $\ket{0_e}$ is the excited-phonon vacuum defined as $b_e \ket{0_e} = 0$. \
  
   Equations~(\ref{3},\ref{b}) that relate $(\hat{x},\hat{p}_x)$ to $(b_g,b_e)$ give the link between these operators as    \begin{subequations} \label{operators}
   \begin{align} 
   \gamma_{_+} b_g+ \gamma_{_-}b_g^\dagger= b_e+\Lambda_e ,  \label{10}  \\
    \gamma_{_+}b_e- \gamma_{_-}b_e^\dagger=  b_g -\Lambda_g  \label{11} , 
  \end{align}
   \end{subequations}
  where the dimensionless prefactors are
     \begin{eqnarray} \label{gamma}
     \gamma_{_\pm}=\frac{1}{2}  \Big(\sqrt {\frac{ \omega_e}{ \omega_g}   }  \pm \sqrt {\frac{ \omega_g}{ \omega_e}   } \Big )
     \, , \,\, \, \, \,  \, \, \, \, 
          \Lambda_{e,g}=l\sqrt {\frac{m\, \omega_{e,g}}{ 2\hbar}   } \, . \quad 
      \end{eqnarray}
      When the two frequencies ($\omega_g, \omega_e$) are equal, $ \gamma_{_+}$ reduces to one, $ \gamma_{_-}$ to zero and $\Lambda_{g}=\Lambda_{e}$.
      Equation (\ref{operators}) allows switching back and forth from ground- and excited-phonons, depending on which basis is more convenient to perform calculation.

    \noindent  \emph{\textbf{Usual independent-boson Hamiltonian}.---}
      By not ing that in a two-level system, $\ket{g} \bra{g}+ \ket{e} \bra{e}$ is the identity operator, we can rewrite $\mathcal{H}_g+\mathcal{H}_e$ as
     \begin{eqnarray}
\mathcal{H} &=&  \Big( \epsilon_g \ket {g} \bra{g}  + \epsilon'_e\ket{e}\bra{e}  \Big) + \Big(  \frac{p_x^2}{2m}  + \frac{m \omega_g^2}{2} x^2 \Big)    \nonumber \\
&& -\Big(   m \omega_e^2 l  x + m \frac{\omega_g^2 - \omega_e^2}{2} x^2 \Big) \ket{e}\bra{e} , 
       \end{eqnarray}
        with $\epsilon_e'=\epsilon_e+ m \omega_e^2 l^2/2 = \epsilon_e + \hbar \omega_e \Lambda_e^2$. As, from Eq.~(\ref{3}), $\hat{x}$ is proportional to $(b_g+b\dg_g)$,
        the above Hamiltonian reads in terms of ground-phonon operators as
\begin{eqnarray} \label{H}
H=H_e+H_g= \epsilon_g  a\dg_g a_g + \epsilon_e' a\dg_e a_e +\hbar \omega_g \big(b_g\dg b_g +1/2\big) 
\nonumber  \\      
 - \hbar \omega_e  \left(\lambda_1 \big(b_g +  b\dg_g\big) +  \lambda_2 \big(b_g +  b\dg_g\big)^2 \right)a\dg_e a_e\, ,  \,\,\,\,\,\,\, 
  \end{eqnarray}
the `coupling' constants being given by $\lambda_1= \Lambda_e\sqrt{ \omega_e / \omega_g}$, and $\lambda_2= (\omega_g^2 - \omega_e^2)/4  \omega_e \omega_g$ which cancels for  $\omega_g =\omega_e$. \

 The usual procedure starting from the Hamiltonian (\ref{H}), is to follow Huang and Rhys \cite{Huang1950a} and use the `polaron' transformation $e^{S}He^{-S}$ with $S= -a\dg_e a_e \, ( b_g\dg -  b_g) \lambda_1$ in the case of `linear' coupling, that is, $\lambda_2=0$ and $\lambda_1=\Lambda_{g}=\Lambda_{e} \equiv\Lambda$. The excited-level eigenstate then appears as \cite{Duke1965a} 
\begin{equation} \label{eq:psi_Duke}
\ket{\Psi_{e,p}^{\rm (pol)}} = e^{a\dg_e a_e \, (b_g\dg- b_g ) \lambda_1}\,\, \frac{b_g^{\dagger \, p}}{\sqrt{p!}} \,a\dg_e \ket{v} \otimes\ket{0_g}, 
\end{equation}
the physical meaning of this state being hard to catch. This state simply is $\ket{e} \otimes \ket{p_e}$  given in Eq.~(\ref{eq:psi_e}). To show it, we first split $e^{(b\dg_g - b_g) \lambda_1} $ as  $e^{-\Lambda^2/2} e^{\Lambda b_g\dg}e^{-\Lambda b_g }$ and note that 
$e^{-\Lambda b_g} \: b_g^{\dagger \, p} \:  e^{\Lambda b_g}  = (b\dg_g - \Lambda)^p = b_e^{\dagger \, p},$   
which leads to
$
|\Psi_{e,p}^{\rm (pol)} \rangle =  \ket{e}\otimes  e^{-\Lambda^2/2}  \frac{b_e^{\dagger \, p}}{\sqrt{p!}} e^{\Lambda b_g\dg}\ket{0_g}.
$
To end the identification, we  write the excited-phonon vacuum $\ket{0_e}$ in terms of ground-phonon states. From its definition, $b_e \ket{0_e} = 0$,  we get $ \ket{0_e}=\sum_p \frac{\Lambda^p}{\sqrt{p!}} u_p \ket{p_g}$ through a recursion relation $u_{p+1}-u_p=0$ that  for $\braket{0_e}{0_e}=1$ yields 
  \begin{eqnarray} \label{eq:tvac}
  \ket{0_e}= u_0 \sum_{p=0}^\infty \frac{\Lambda^p }{\sqrt{p!} } \ket{p_g}=
   \frac{e^{ \Lambda b_g\dg}}{e^{\Lambda^2 /2}} \ket{0_g} , 
  \end{eqnarray}
  from which it is easy to see that $|\Psi_{e,p}^{\rm (pol)} \rangle$ is indeed equal to $\ket{e} \otimes \ket{p_e}$.

 The above excited-phonon vacuum $\ket{0_e}$ has some interesting insights: It is made of states having an arbitrarily large number of ground-phonons. Yet, the number of ground-phonons it contains is finite,
  \begin{eqnarray}
  \label{eq:Lambda}
  \bra{0_e}b_g\dg b_g \ket{0_e}=   \Lambda ^2 , \
   \end{eqnarray}  
which directly follows from Eq.~(\ref{11}). So, the square of the interaction parameter $ \lambda_1$, known as the Huang-Rhys's factor, just corresponds to the ground-phonon number in the excited-phonon vacuum.  
 The ground-phonon vacuum energy, $\bra{0_e}\hbar \omega b_g\dg b_g \ket{0_e}=\hbar \omega \Lambda^2$, is commonly known as the `reorganization energy', and often added to the exited level through a modified energy $\epsilon_e'$ as in Eq. (\ref{H}).

\noindent \emph{\textbf{Effect of spatial shift}.---} 
Let us first focus on linear coupling, $\omega_e=\omega_g\equiv\omega$ and study the time evolution of the initial state $\ket{e} \otimes \ket{p_g}$, obtained from a sudden 
photon absorption in the ground-level eigenstate $\ket{\psi_{p,g}} = \ket{g} \otimes \ket{p_g}$, namely 
\begin{eqnarray} \label{eq:psi_et}
\ket{\phi_{e,p;t}} {=} 
e^{-\frac{i}{\hbar} H_e t} \ket{e} \otimes \ket{p_g}   {=}
 e^{- \frac{i}{\hbar} \epsilon_e  t} \ket{e} \otimes \ket{p_{g;t}} 
\end{eqnarray}
with 
$
\ket{p_{g;t}}= e^{- i \omega t (b_e\dg b_e + 1/2)} \ket{p_g} .
$

The best way to perform calculations is to switch back and forth from ground-phonons to excited-phonons using Eq.~(\ref{operators}), and to use 
$b \,e^{-i\omega t b\dg b} {= } e^{-i\omega t( b\dg b+1)}\, b $, valid for $b=(b_g,b_e)$ since  $ \big[b_e, b_e\dg\big]{=}1{=}\big[b_g,b_g\dg\big]$. This leads to
\begin{equation}
b_g e^{- i \omega t b_e\dg b_e} =e^{- i \omega t (b_e\dg b_e+1)} \big( b_g{-}\Lambda_t \big),  \label{bexpm}\\ 
\end{equation}
the time-dependent linear coupling being 
\begin{equation} \label{22'}
\Lambda_t= \Lambda (1-e^{i \omega t}).
\end{equation} 
This readily
gives the number of ground-phonons in the $\ket{\phi_{e,p;t}}$ state as
\begin{equation} \label{Nt}
\bra{p_{g;t}} b_g\dg b_g \ket{p_{g;t}} = p + |\Lambda_t|^2=p + 4\Lambda^2 \sin^2(\omega t /2) .  
\end{equation}
It oscillates between its initial value, $p$, and $p+4\Lambda^2$. By contrast, the number of excited-phonons in this state stays constant, $\bra{p_{g;t}} b_e\dg b_e \ket{p_{g;t}}=p+ \Lambda^2$, making the Hamiltonian mean value of the $\ket{\phi_{e,p;t}}$ state,
$ \bra{\phi_{e,p;t}} H_e  \ket{\phi_{e,p;t}}$, constant and equal to $\epsilon_e + \hbar \omega (p+ \Lambda^2 + 1/2) $, in spite of the oscillation of the ground-phonon number. This is an additional supportive point that ground-phonons have no physical relevance when the electronic system is in its excited level.\

 Another important quantity is the time-dependent correlation function for spectral lineshape. 
For a two-level system in a phonon bath at thermal equilibrium with the system ground state, the absorption lineshape obtained from the Fermi golden rule follows from 
\begin{equation}\label{eq:Gt}
G_{g;t} =  \sum_{p=0}^\infty \frac{e^{- \beta\hbar \omega p }}{Z} G_{g,p;t}
\end{equation}
where $T = 1/k_B \beta$ is the temperature and $Z=\sum_p e^{- \beta\hbar \omega p } = (1 - e^{-\beta\hbar \omega})^{-1}$ is the phonon partition function.  $G_{g,p;t} $ is the correlation function associated with the transition dipole moment $\mu = a\dg_e a_g + h.c.$, namely
$G_{g,p;t} = \bra{\psi_{g,p}} \mu_t \, \mu_0 \ket{\psi_{g,p}}$ 
with $\mu_t= e^{\frac{i}{\hbar} H t}   \mu   e^{-\frac{i}{\hbar} H t} $. By separating the electronic part from the phonon part,  $G_{g,p;t}$ reduces to 
$ e^{- i t \Omega_{eg} } e^{ i \omega t (p+ 1/2)}   \braket{p_g}{p_{g;t}}$,
the  frequency $\Omega_{eg} $ corresponding to the electronic transition, $(\epsilon_e - \epsilon_g )/ \hbar$. 
We get the overlap through 
$
\braket{p_g}{p_{g;t}}\equiv e^{-i \omega t (p+1/2) }T_p
$, with
\begin{eqnarray}\label{24"}
T_p = \bra{p_g} e^{-i \omega t (b_e\dg b_e-p) } \ket{p_g}.
\end{eqnarray}
This quantity, equal to 1 for $b_e= b_g$, fulfills   
\begin{equation}
\label{22}
0=p T_p -\Big (2p - 1 - |\Lambda_t|^2\Big) T_{p-1} + (p-1) T_{p-2} , 
\end{equation}
which is the recursion relation for Laguerre polynomials. 
So, $T_p = T_0 L_p(|\Lambda_t|^2)$, with $ T_0  = e^{-\Lambda \Lambda^{*}_t}$ obtained from $dT_0/dt$. All this yields   
\begin{equation} \label{overlap_mm}
\braket{p_g}{p_{g;t}} = e^{-\Lambda \Lambda^{*}_t} e^{-i \omega t  (p + 1/2) } L_p(|\Lambda_t|^2) . 
\end{equation}

The thermal average in (\ref{eq:Gt}) is easy to perform  by using $\sum_{p=0}^\infty z^p L_p(x) = e^{-z x / (1-z)}/(1-z)$. This readily gives the established result in terms of the average phonon number $\bar{p} = \sum_{p} p e^{-\beta\hbar \omega p } = (e^{\beta\hbar \omega} - 1)^{-1}$, namely 
\begin{eqnarray} \label{Gt}
G_{g;t} &{=}&  e^{-i \Omega_{eg} t}  e^{-\Lambda \Lambda_t^*}  (1 {-} e^{-\beta\hbar \omega})  { \sum_{p=0}^\infty} e^{- \beta\hbar \omega p }  L_p(|\Lambda_t|^2) \nonumber \\
&{=}&  e^{- i  \Omega_{eg}t } e^{- \Lambda \Lambda_t^*} e^{- \bar{p} |\Lambda_t|^2}, 
\end{eqnarray}
as  derived in far heavier ways using a combination of  polaron transformation, Green's function and Feynman's disentanglement \cite{MahanBook}, or   a combination of  interaction picture,  cumulant expansion and Wick's theorem  \cite{MukamelBook}. 
 The absorption spectrum then follows from $A(w) = 2 {\rm Re} (\int_0^\infty dt \, e^{i w t} G_{g;t})$, its zero-temperature amplitude at the pole $\Omega_{eg} + n\omega$ being the Poisson distribution $2\pi\Lambda^{2n}e^{-\Lambda^2}/n!$.  
When the excited state is at thermal equilibrium with the phonon bath, the emission spectrum can be likewise obtained. 

\begin{figure*}
\includegraphics[width=2\columnwidth]{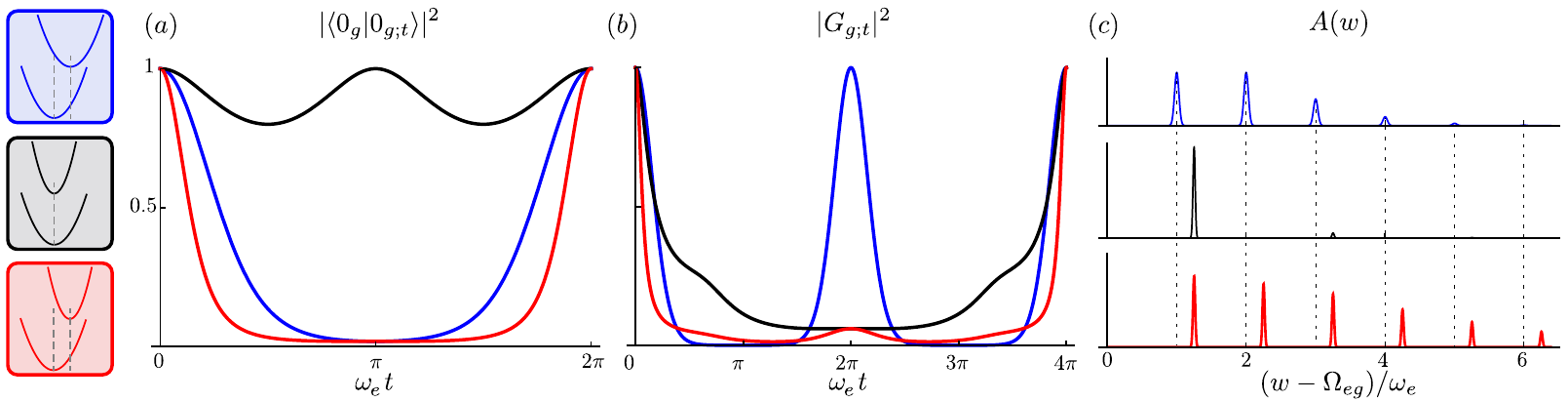}
\caption{Comparison between three phonon configurations sketched on the left, from top to bottom: (i) Linear coupling ($\omega_g = \omega_e, l\neq0$), (ii) quadratic coupling ($\omega_e \neq \omega_g, l=0)$, and (iii) both $ ( \omega_e \neq \omega_g, l\neq0)$. We show (a) the time evolution of $|\braket{p_g}{p_{g;t}}|^2$ given in Eqs.~(\ref{overlap_mm}, \ref{37''}) when the initial state is the ground-phonon vacuum, $p=0$;  (b) the correlation function $|G_{g;t}|^2$ given in Eqs.~(\ref{Gt}, \ref{gt}), for $\beta \hbar \omega_e=1$; and (c) the absorption spectrum $A(w)$ given in Eq.~(\ref{Ageneral}) for $T=0$, the delta functions being taken as Gaussians for representation purpose.  
 (i) Blue curves: Linear coupling only ($\omega_g = \omega_e \equiv \omega$ and $1=\Lambda_g=\Lambda_e \equiv\Lambda$); (ii) black curves: Quadratic coupling only ($\omega_e =2 \omega_g$ and $0=\Lambda_g=\Lambda_e)$; (iii) red curves: Both couplings $(\omega_e =2 \omega_g$ and $1=\Lambda_g=\Lambda_e/\sqrt{2})$.   
\label{fig2}}
\end{figure*}

\noindent \emph{\textbf{Spatial shift and frequency change}.---} 
 Ground- and excited-phonons prove even more useful for different phonon frequencies, $\omega_g\neq \omega_e$, that is, when the quadratic term is present in Eq. (\ref{H}). In this case, it is  still possible to find analytical expressions for the quantities considered in the linear limit, by using the same commutation procedure. To catch the consequences of different phonon frequencies, we  here only give new expressions of some relevant quantities, with a few hints on how to obtain them. Detailed derivations will be given in an extended version. Some  consequences of having a spatial shift, a frequency change or both are visualized in Fig.~\ref{fig2}.

When $\omega_e\neq \omega_g$, that is, $\gamma_{_-}\neq0$, the operator $b_g$ given in Eq.~(\ref{11}) not only destroys but also creates an excited-phonon. Compared to Eq.~(\ref{eq:Lambda}), this leads to an increase of the number of ground-phonons in the excited-phonon vacuum,  
\begin{eqnarray} 
\bra{0_e} b_g\dg b_g \ket{0_e} = \Lambda_g^2 + \gamma_{_-}^2,  
\end{eqnarray}
This also transforms Eq.~(\ref{bexpm}) into
\begin{equation}\label{bexpm3}
  b_g e^{- i \omega_e t b_e\dg b_e} = e^{- i \omega_e t (b_e\dg b_e+1)} \big(\tilde{D}'_t b_g +\tilde{Q}'_t  b\dg_g{-}\tilde{\Lambda}'_t        \big) ,
\end{equation}
with $\tilde{D}'_t= \gamma_{_+}^2 - \gamma_{_-}^2 e^{2i \omega_e t}$, $\tilde{Q}'_t= \gamma_{_+}\gamma_{_-} (1 - e^{2i \omega_e t})$, and 
$
\tilde{\Lambda}'_t = \Lambda_e (\gamma_{_+}  - \gamma_{_-} e^{2i \omega_e t}) - \Lambda_g e^{i \omega_e t}
$.
The existence of two time-dependent terms, $e^{i \omega_e t}$ and $e^{2i \omega_e t}$, instead of $e^{ i \omega t}$ only  in the linear limit (see Eq.(\ref{22'})), brings two oscillatory terms in the time evolution of the ground-phonon number, which now reads 
\begin{flalign} \label{eq:N}
&\bra{p_{g;t}} b_g\dg b_g \ket{p_{g;t }}= p |\tilde{D}'_t|^2 + (p+1) |\tilde{Q}'_t|^2 + |\tilde{\Lambda}'_t |^2 &
\\
&=p{+}4\Lambda_g^2 \sin^2\frac{\omega_e t}  {2}
  {+}\frac{\omega_e ^2 {-}\omega_g ^2}  {\omega_e\omega_g}\Big (\Lambda_e^2{+}(2p{+}1)\frac{\omega_e ^2 {-}\omega_g ^2}  {4 \omega_e\omega_g} \Big)  \sin^2 \omega_e t  .&
   \nonumber
   \end{flalign}
  The quadratic coupling adds a faster oscillation to the time evolution of the ground-phonon number, with an amplitude  that depends on the phonon number $p$. This faster oscillation, which disappears when $\omega_e=\omega_g$ in agreement with  Eq.~(\ref{Nt}), still exists for zero spatial shift, $l=0$, that is, in the absence of linear coupling, $\Lambda_e=\Lambda_g=0$.

The correlation function for the spectral lineshapes requires the knowledge of  $T_p$ defined in Eq.~(\ref{24"}) with $\omega$ taken equal to $\omega_e$. The fact that the operator $b_g$ not only destroys but also creates an excited-phonon brings a recursion relation between four $T_p$'s, instead of three as in Eq.(\ref{22}). For $T_p\equiv \tilde{D}_t^{'p} \tilde{T}_p$, this recursion relation reads
\begin{flalign}
\label{34'}
&0=a_0 \, p \tilde{T}_p-\Big((p-1) a_1+b_1\Big)\tilde{T}_{p-1}&
 \\
&\quad +\Big((p-2) a_2+b_2\Big)\tilde{T}_{p-2} -\Big((p-3) a_3+b_3\Big)\tilde{T}_{p-3} 
\nonumber
\end{flalign}
for $p\ge 3$ while for $p=(1,2)$, it reduces to its first two and three terms, respectively. In the linear limit, this equation reduces to the difference of the recursion relations (\ref{22}) for $T_p$ and $T_{p-1}$; so, we do recover the `linear-coupling' result, but in a non-trivial way. \

To solve the above recursion relation (\ref{34'}) beyond the linear limit, we introduce the generating function $\tilde{K}(x)= \sum_{p=0}^\infty  x^p \tilde{T}_p$. It obeys a first-order differential equation
\begin{equation}
\label{32'}
0{=}\Big(a_0{-}a_1x{+}a_2x^2{-}a_3x^3\Big)
 \frac{ d\tilde{K}(x)} {dx}    
 {+}\Big(-b_1{+}b_2x{-}b_3x^2\Big)\tilde{K}(x)\: 
\end{equation}
the solution of which reads, by using the explicit  $(a_i,b_i)$ values, as 
  \begin{eqnarray}\label{36''}
\tilde{K}(x)= 
\tilde{T}_0{\sqrt \frac{1-\tilde{Q}^{2}_t}{(x{-}1)^2- \tilde{Q}_t^2}}
\exp \Big(  \frac{-  x \,\,\tilde{\Lambda}^{2}_t/ \tilde{D}_t}{(x-1+\tilde{Q}_t )(1-\tilde{Q}_t ) }\Big){} \,\,\,
   \end{eqnarray}
   with $ \tilde{Q}_t=e^{-i  \omega_e t}   \tilde{Q}'_t$ and similarly for $( \tilde{D}_t, \tilde{\Lambda}_t)$.

To get $T_p$, we must extract the $x^p$ coefficient of $\tilde{K}(x)$. We do it by rewriting $\tilde{K}(x)$ as a product of two functions that are easy to expand in powers of $x$. After some algebra, we end with
\begin{eqnarray} \label{37''}
\braket{p_g}{p_{g;t}}&=& \tilde{T}_0 \: e^{-  i \omega_e t/2} \left(\frac{\tilde{D}_t}{1+\tilde{Q}_t} \right)^p \\
 & \times & \sum_{k=0}^p    \left( \frac{1+\tilde{Q}_t }{1-\tilde{Q}_t }\right)^k L^{-\frac{1}{2}}_{p-k}(0) L^{-\frac{1}{2}}_{k}\Big(\frac{ -\tilde{\Lambda}^{2}_t}{\tilde{D}_t(1-\tilde{Q}_t)}\Big)
  \nonumber
\end{eqnarray}
with $\tilde{T}_0$ given by 
 \begin{equation} \label{T0'}
   \tilde{T}_0 =( \gamma_{_+}^2 - \gamma_{_-}^2 e^{-2 i \omega_e t} )^ {-1/2}
  \exp({-\Lambda_g \Lambda_e \dfrac{1-e^{- i \omega_e t}}{ \gamma_{_+} - \gamma_{_-} e^{- i \omega_e t}}} ) .
   \end{equation}
This result reduces to Eq.~(\ref{overlap_mm}) in the linear limit: Indeed, $\tilde{Q}_t=0$ and  $ -\tilde{\Lambda}^{2}_t / \tilde{D}_t=|\Lambda_t|^2$ while the sum in Eq.~(\ref{37''}) follows from the addition formula of Laguerre polynomials \cite{AbramowitzBook}, i.e.,  
$
\sum_{k=0}^p   L^{-\frac{1}{2}}_{p-k}(0)   L_p^{-\frac{1}{2}}(|\Lambda_t|^2) = L_p(|\Lambda_t|^2). 
$

\noindent \emph{\textbf{Consequences of spatial shift and frequency change}.---} 
Figure~\ref{fig2}a shows $|\braket{p_g}{p_{g;t}}|^2$ for the three relevant coupling configurations, namely, a spatial shift $l\neq0$, a frequency change $\omega_g\neq\omega_e$ and both, when the initial state is the ground-phonon vacuum.  We see that the likelihood that the phonon number state remains in the initial state is much smaller when $l\neq0$ than when $l=0$, that is, when the linear coupling vanishes. Mathematically, this is due to the fact that when $\Lambda_{e}=\Lambda_{g}=0$, the exponential factor in  $\tilde{T}_0$ reduces to 1. We also find that, for finite $\Lambda_{g,e}$'s, the likelihood for the phonon state to remain in its initial state is smaller when $\omega_g\neq\omega_e$ than when $\omega_g=\omega_e$, that is, when the quadratic coupling vanishes.
These observations are born out by the changing distribution of the absorption spectra, shown in Fig.~\ref{fig2}c.

With $\braket{p_g}{p_{g;t}}$ known, it becomes possible to derive  the correlation function for the absorption lineshape when the phonon bath is at thermal equilibrium, as defined in Eq.~(\ref{eq:Gt}) with $\omega$ taken as $\omega_g$. It reads   
\begin{eqnarray}
G_{g;t}=\frac{1}{Z_g} e^{- i \Omega_{eg}t}  e^{  i (\omega_g{-}\omega_e) \frac{t}{2}}  \sum_{p=0}^\infty (e^{-\beta \hbar \omega_g + i \omega_g t} \tilde{D}_t)^p \tilde{T}_p, \quad \:\:
\end{eqnarray}
where $Z_g$ is the partition function calculated with $\omega_g$. 
This quantity can be conveniently written in terms of the generating function $\tilde{K}(x)$ given in Eq.~(\ref{36''}), as 
\begin{flalign} \label{gt}
&G_{g;t}=\frac{1}{Z_g} e^{- i  \Omega_{eg} t} e^{  i (\omega_g-\omega_e) t/2}  & \\
& \hspace*{5em}\times  \tilde{K}\Big(e^{-\beta \hbar \omega_g} (\gamma_{_+}^2 e^{i (\omega_g-\omega_e) t} - \gamma_{_-}^2 e^{i (\omega_g+ \omega_e) t})\Big)  . \nonumber
\end{flalign}
For $\gamma_{_-} = 0$, it reduces to the linear limit, Eq.~(\ref{Gt}), while for zero temperature, $G_{g;t}$ reduces to $e^{- i  \Omega_{eg} t} e^{  i (\omega_g-\omega_e) t/2}  \tilde{T}_0 $. By writing $\tilde{T}_0$ in terms of Hermite polynomials ${\rm H}_n(x)$ through the Mehler's formula, we can get the absorption spectrum at $T=0$ as  illustrated in Fig.~\ref{fig2}c and given by 
\begin{eqnarray}\label{Ageneral}
A(w)\!\! &=&\!\! \frac{2\pi}{\gamma_+} e^{- \frac{\Lambda_e \Lambda_g}{\gamma_+}} \sum_{n=0}^{\infty} \frac{1}{n!}\! \left( - \frac{\gamma_- }{2\gamma_+} \right)^n \left( {\rm H}_n\Big(\frac{\Lambda_g}{\sqrt{-2\gamma_+ \gamma_-}}\Big)\right)^2 \nonumber\\
 &&\hspace{1cm}\times \delta\big(w - \Omega_{eg} - \frac{\omega_e - \omega_g}{2} - n \omega_e\big) .
\end{eqnarray}


The above results (\ref{gt},\ref{Ageneral}), together with Eqs.~(\ref{37''},\ref{T0'}), are the \textbf{\emph{first analytical expressions}} that include both `linear' and quadratic couplings for a two-level system coupled to phonons. The correlation function $G_{g;t}$ is shown in Fig.~\ref{fig2}b: Compared to the overlap $\braket{p_g}{p_{g;t}}$, a change in periodicity occurs when $\omega_g\not=\omega_e$, due to the additional oscillatory terms $e^{i (\omega_g - \omega_e)t}$ and $e^{i (\omega_g + \omega_e)t}$ appearing in Eq.~(\ref{gt}). Difference in the phonon frequencies also modifies the position and amplitude of the energy poles in the absorption spectrum. Even at $T=0$, the zero-point-energy difference shifts the spectrum compared to linear coupling (see Fig.~\ref{fig2}c).  Increasing the temperature will add more energy poles to the spectrum as more ground-phonon Fock states come into play.

{\bf \textit{In conclusion,}} we propose a conceptually different approach to the independent boson model. It relies on  a diagonal representation using  two sets of phonons  physically relevant for the problem, namely, ground-phonons and excited-phonons that depend on the level occupied by the electronic system.  By capturing the essence of the problem, this representation removes couplings in a natural way from the very first line, without the need of any transformation. It considerably simplifies the resolution of the problem, by solely relying on commutation relations between the two types of phonon operators. Besides recovering the known results in a simple way, we are able to go further in the model complexity by solving the problem for different phonon frequencies, that is, quadratic coupling, and to study its intricate consequences on the time dependence of the correlation functions for all coupling configurations.  While the independent boson model is the simplest model to study electron-phonon interactions, we anticipate that the physical approach presented here can allow for simpler resolutions of more complex problems, starting with the spin-boson model.

\begin{acknowledgments}
 M.C. wishes to thank Avadh Saxena for invitation at Los Alamos National Laboratory and  Alex Chin for stimulating discussions. A.C. thanks the Institut des NanoSciences de Paris and the DIM SIRTEQ for invitation.  
\end{acknowledgments}


%

\end{document}